\documentclass[preprint,12pt]{elsarticle}

\usepackage{graphicx}
\usepackage{amssymb}
\usepackage{lineno}

\journal{Chaos Solitons \& Fractals}

\begin{document}

\begin{frontmatter}


\title{Heterogeneous diffusion in comb and fractal grid structures}



\author{Trifce Sandev}

\address{Max Planck Institute for the Physics of Complex Systems, N\"{o}thnitzer Strasse 38, 01187 Dresden, Germany\\Radiation Safety Directorate, Partizanski odredi 143, P.O. Box 22, 1020 Skopje, Macedonia\\Research Center for Computer Science and Information Technologies, Macedonian Academy of Sciences and Arts, Bul. Krste Misirkov 2, 1000 Skopje, Macedonia}

\author{Alexander Schulz}

\address{Max Planck Institute for the Physics of Complex Systems, N\"{o}thnitzer Strasse 38, 01187 Dresden, Germany}

\author{Holger Kantz}

\address{Max Planck Institute for the Physics of Complex Systems, N\"{o}thnitzer Strasse 38, 01187 Dresden, Germany}

\author{Alexander Iomin}

\address{Department of Physics, Technion, Haifa 32000, Israel}

\begin{abstract}
We give an exact analytical results for diffusion with a power-law position dependent diffusion coefficient along the main channel (backbone) on a comb and grid comb structures. For the mean square displacement along the backbone of the comb we obtain behavior $\langle x^2(t)\rangle\sim t^{1/(2-\alpha)}$, where $\alpha$ is the power-law exponent of the position dependent diffusion coefficient $D(x)\sim |x|^{\alpha}$. Depending on the value of $\alpha$ we observe different regimes, from anomalous subdiffusion, superdiffusion, and hyperdiffusion. For the case of the fractal grid we observe the mean square displacement, which depends on the fractal dimension of the structure of the backbones, i.e., $\langle x^2(t)\rangle\sim t^{(1+\nu)/(2-\alpha)}$, where $0<\nu<1$ is the fractal dimension of the backbones structure. The reduced probability distribution functions for both cases are obtained by help of the Fox $H$-functions.
\end{abstract}


\end{frontmatter}

%\linenumbers

\section{Introduction}\label{intro}

In many physical systems, such as transport in inhomogeneous media and plasmas \cite{plasma}, and diffusion on random fractals \cite{fractals}, the diffusion coefficient is not a constant but depends on the particle position, like in turbulent diffusion \cite{GP84,HP84}, including  turbulent two-particle diffusion \cite{turbulent}. The heterogeneous diffusion equation has been investigated within the continuous time random walk (CTRW) theory in \cite{BI04,IB05,sro kam}, and the mean first passage time of such systems was analyzed in \cite{fa lenzi}. L\'evy processes in inhomogeneous media \cite{srokowski} and ergodicity breaking in heterogeneous diffusion processes \cite{metzler njp} have been investigated, as well as the influence of external potentials on heterogeneous diffusion processes was recently considered in \cite{rk}. Time and power-law position dependent diffusion coefficient were also considered in the literature \cite{lenzi pla} in analysis of $N$-dimensional diffusion equation. 

The displacement $x(t)$ of a particle in a heterogeneous medium with space dependent diffusivity $\mathcal{D}(x)$ is described by the Langevin equation 
\begin{eqnarray}\label{LEq}
\frac{d}{dt}x(t)=\sqrt{2\mathcal{D}(x)}\zeta(t),
\end{eqnarray}
where $\zeta(t)$ is a white Gaussian noise with $\left\langle\zeta(t)\zeta(t')\right\rangle=\delta(t-t')$ and zero mean $\left\langle\zeta(t)\right\rangle=0$. In the Stratonovich interpretation this Langevin equation corresponds to the diffusion equation for the probability distribution function (PDF) \cite{metzler njp}
\begin{eqnarray}\label{heterogeneous_diffusion} 
\frac{\partial}{\partial t}P(x,t)
=\frac{\partial}{\partial x}\left[\sqrt{\mathcal{D}(x)}\frac{\partial}{\partial x}\left(\sqrt{\mathcal{D}(x)}P(x,t)\right)\right].
\end{eqnarray}
It is supplemented with the initial condition $P(x,t=0)=\delta(x)$, and the boundary conditions are set to zero at infinities. The diffusion coefficient has the power-law position dependent form 
\begin{eqnarray}\label{D(x)}
\mathcal{D}(x)=\mathcal{D}_{x}|x|^{\alpha}\, , \quad \quad \alpha<2\, .
\end{eqnarray}
The solution of Eq.~(\ref{heterogeneous_diffusion}) is obtained in the stretched exponential form \cite{metzler njp}
\begin{eqnarray}\label{sol het diff}
P(x,t)=\frac{|x|^{-\alpha/2}}{\sqrt{4\pi\mathcal{D}_{x}t}}\exp\left(-\frac{|x|^{2-\alpha}}{(2-\alpha)^{2}\mathcal{D}_{x}t}\right),
\end{eqnarray}
and the mean square displacement (MSD) has the power-law dependence on time
\begin{eqnarray}\label{MSD het diff}
\left\langle x^{2}(t)\right\rangle=\int_{-\infty}^{\infty}dx\,x^{2}P(x,t)\simeq \frac{t^{\frac{2}{2-\alpha}}}{\Gamma\left(1+\frac{2}{2-\alpha}\right)}.
\end{eqnarray}
This expression describes different diffusiove regimes, where for $\alpha<0$ one observes subdiffusion, normal diffusion for $\alpha=0$, superdiffusion for $0<\alpha<1$, ballistic motion for $\alpha=1$ and hyperdiffusion for $1<\alpha<2$. The case with $\alpha=2$ leads to exponentially fast spreading \cite{BI04,iomin pre2013}. The case with $\alpha>2$ yields localization\footnote{In Ref. \cite{IB05}, where inhomogeneous advection in a comb was considered, this regime has been named by negative superdiffusion.} with the decay MSD $~t^{-\frac{2}{\alpha-2}}$.

A random walk in a simple comb structure consisting of main diffusion channel (backbone) and trapping fingers leads to anomalous diffusion with a transport exponent equal to $1/2$ \cite{comb ex}. It can be described by the two-dimensional diffusion equation \cite{ark}
\begin{equation}\label{diffusion eq on a comb}
\frac{\partial}{\partial t}P(x,y,t)
=\mathcal{D}_{x}\delta(y)\frac{\partial^{2}}{\partial x^{2}}P(x,y,t)+\mathcal{D}_{y}\frac{\partial^{2}}{\partial y^{2}}P(x,y,t),
\end{equation}
where $P(x,y,t)$ is the probability distribution function (PDF), $\mathcal{D}_{x}\delta(y)$ is the diffusion coefficient in $x$ direction with dimension
$[\mathcal{D}_{x}]=\mathrm{m}^{3}/\mathrm{s}$, and
$\mathcal{D}_{y}$ is the diffusion coefficient in $y$ direction with dimension
$[\mathcal{D}_{y}]=\mathrm{m}^{2}/\mathrm{s}$. The
$\delta$-function in Eq.~(\ref{diffusion eq on a comb}) means that the diffusion along the $x$ direction occurs only at $y=0$ (the backbone) and the fingers play the role of traps. The comb model (\ref{diffusion eq on a comb}) is used to describe diffusion in low-dimensional percolation clusters \cite{ark,comb application}. Comb models can further be generalized to grid and fractal grid structures \cite{fractal grid pre} in which the diffusion along the $x$ direction may appear in many backbones, even infinite number of backbones which positions belong to a fractal set $\mathcal{S}_{\nu}$ with fractal dimension $0<\nu<1$. In this case anomalous diffusion is observed and the transport exponent depends on the fractal dimension $\nu$. In this paper we consider heterogeneous diffusion on such comb and fractal grid structures, where the diffusivity is position dependent with power-law diffusion coefficient of Eq. (\ref{D(x)}).

The investigation of anomalous diffusion processes in complex systems leads to appearance of fractional differintegration in the corresponding stochastic and kinetic equation representing the memory effect in the system. Therefore, the mathematical background of the theory of fractional differential and integral equations \cite{SKM book,Podlubny,yz}, and associated Mittag-Leffler and Fox $H$-functions \cite{erdelyi,saxena book} for analysis of such processes are of the primary importance. From the other side, diffusion on fractal structures, and the connection between the fractal dimension and fractional differintegration, as well as description of fractal processes by fractional calculus have been discussed in the scientific community \cite{west}.

The paper is organized as follows. In Sec. \ref{sec2} we consider a two-dimensional diffusion equation for a comb with the position dependent (power-law) diffusion coefficient along the backbone. Exact results for the PDF and MSD are obtained and various diffusion regimes are observed, such as anomalous subdiffusion, superdiffusion and hyperdiffusion. The case of heterogeneous diffusion on a fractal grid structure is considered in Sec. \ref{sec3}, and exact results for the PDF and MSD are derived. The summary is given in Sec. \ref{sec4}.

\section{Heterogeneous diffusion on a comb}\label{sec2}

We consider the following heterogeneous two dimensional diffusion equation on a comb for the PDF $P(x,y,t)$
\begin{eqnarray}\label{heterogeneous_comb} 
\frac{\partial}{\partial t}P(x,y,t)
&=&\delta(y)\frac{\partial}{\partial x}\left[\sqrt{\mathcal{D}(x)}\frac{\partial}{\partial x}\left(\sqrt{\mathcal{D}(x)}P(x,y,t)\right)\right]\nonumber\\&+&\mathcal{D}_{y}\frac{\partial^{2}}{\partial y^{2}}P(x,y,t),
\end{eqnarray}
where $\mathcal{D}(x)$ is the position dependent diffusion coefficient along the backbone, $\mathcal{D}_{y}$ is the diffusion coefficient along the fingers. This equation is a generalization of the one-dimensional heterogeneous diffusion equation (\ref{heterogeneous_diffusion}) to a two-dimensional comb structure. The initial condition is
\begin{equation}\label{initial condition}
P(x,y,t=0)=\delta(x)\delta(y),
\end{equation}
and the boundary conditions for $P(x,y,t)$ and $\frac{\partial}{\partial q}P(x,y,t)$, $q=\{x,y\}$ are set to zero at infinities, $x=\pm\infty$, $y=\pm\infty$. The position dependent diffusion coefficient has power-law form (\ref{D(x)}) with $\alpha<2$, therefore the physical dimension of the diffusion coefficient along the backbone $\mathcal{D}_{x}\delta(y)$ is $\left[\mathcal{D}_{x}\delta(y)\right]=\mathrm{m}^{2-\alpha}\mathrm{s}^{-1}$, and the physical dimension of $\mathcal{D}_{y}$ is $\left[\mathcal{D}_{y}\right]=\mathrm{m}^{2}\mathrm{s}^{-1}$.

Inserting the diffusion coefficient (\ref{D(x)}) in 
Eq.~(\ref{heterogeneous_comb}) one obtains
\begin{eqnarray}\label{heterogeneous_comb power-law} 
\frac{\partial}{\partial t}P(x,y,t)
=\mathcal{D}_{x}\delta(y)\frac{\partial}{\partial x}\left[|x|^{\alpha/2}\frac{\partial}{\partial x}\left(|x|^{\alpha/2}P(x,y,t)\right)\right]+\mathcal{D}_{y}\frac{\partial^{2}}{\partial y^{2}}P(x,y,t).
\end{eqnarray}

From the Laplace transform\footnote{The Laplace transform of a given function $f(t)$ is defined by $f(s)=\mathcal{L}[f(t)]=\int_{0}^{\infty}dt\,e^{-st}f(t)$.}, it follows
\begin{eqnarray}\label{heterogeneous_comb_L} 
sP(x,y,s)-P(x,y,t=0)
&=&\mathcal{D}_{x}\delta(y)\frac{\partial}{\partial x}\left[|x|^{\alpha/2}\frac{\partial}{\partial x}\left(|x|^{\alpha/2}P(x,y,s)\right)\right]\nonumber\\&+&\mathcal{D}_{y}\frac{\partial^{2}}{\partial y^{2}}P(x,y,s).
\end{eqnarray}
We present the solution of the Eq.~(\ref{heterogeneous_comb_L}) in the form of the ansatz
\begin{eqnarray}\label{ansatz}
P(x,y,s)=g(x,s)\exp\left(-\sqrt{\frac{s}{\mathcal{D}_{y}}}|y|\right),
\end{eqnarray}
from where it follows that
\begin{eqnarray}\label{ansatz y=0}
P(x,y=0,s)=g(x,s)\, .
\end{eqnarray}
We also introduce the reduced PDF, which describes the transport along the backbones only 
$$p_{1}(x,t)=\int_{-\infty}^{\infty}dy\,P(x,y,t),$$ 
and yields
\begin{eqnarray}\label{p1}
p_{1}(x,s)=2g(x,s)\sqrt{\frac{\mathcal{D}_{y}}{s}}.
\end{eqnarray}
Integrating Eq.~(\ref{heterogeneous_comb power-law}) over $y$, one finds
\begin{eqnarray}\label{heterogeneous_comb power-law p1} 
sp_{1}(x,s)-p_{1}(x,t=0)=\mathcal{D}_{x}\frac{\partial}{\partial x}\left[|x|^{\alpha/2}\frac{\partial}{\partial x}\left(|x|^{\alpha/2}g(x,s)\right)\right],
\end{eqnarray}
where the initial condition $p_1(x,t=0)=\delta(x)$. Therefore, from Eqs.~(\ref{heterogeneous_comb power-law p1}) and (\ref{p1}) we obtain the differential equation
\begin{eqnarray}\label{heterogeneous_comb power-law p1 final} 
2\sqrt{\mathcal{D}_{y}}s^{1/2}g(x,s)-\mathcal{D}_{x}\frac{\partial}{\partial x}\left[|x|^{\alpha/2}\frac{\partial}{\partial x}\left(|x|^{\alpha/2}g(x,s)\right)\right]=\delta(x).
\end{eqnarray}
After the substitution $f(x,s)=|x|^{\alpha/2}g(x,s)$, from Eq.~(\ref{heterogeneous_comb power-law p1 final}) we obtain
\begin{eqnarray}\label{heterogeneous_comb power-law f final} 
2\sqrt{\mathcal{D}_{y}}s^{1/2}|x|^{-\alpha/2}f(x,s)-\mathcal{D}_{x}\frac{\partial}{\partial x}\left[|x|^{\alpha/2}\frac{\partial}{\partial x}f(x,s)\right]=\delta(x).
\end{eqnarray}
We take into account symmetrical property of the equation, which is invariant with respect to inversion $x\rightarrow  -x$. Therefore, in order to solve this equation, we use $z=|x|$, from where by partial differentiation with respect to $x$ we find\footnote{We also use here the following property $x = |x|{\rm sign}(x)$, and 
${\rm sign}(x)\partial_z =\partial_x$.}
\begin{eqnarray}\label{heterogeneous_comb power-law f(y) final} 
&&2\sqrt{\mathcal{D}_{y}}s^{1/2}z^{-\alpha/2}f(z,s)-\mathcal{D}_{x}(\alpha/2)z^{\alpha/2-1}\frac{\partial}{\partial z}f(z,s)-\mathcal{D}_{x}z^{\alpha/2}\frac{\partial^{2}}{\partial z^{2}}f(z,s)\nonumber\\&&-2\mathcal{D}_{x}z^{\alpha/2}\frac{\partial}{\partial z}f(z,s)\delta(x)=\delta(x).
\end{eqnarray}
This equation splits into the system of equations
\begin{equation}\label{heterogeneous_comb power-law f(y) final homogeneous part} 
2\sqrt{\mathcal{D}_{y}}s^{1/2}z^{-\alpha/2}f(z,s)-\mathcal{D}_{x}(\alpha/2)z^{\alpha/2-1}\frac{\partial}{\partial z}f(z,s) %\nonumber\\&&
-\mathcal{D}_{x}z^{\alpha/2}\frac{\partial^{2}}{\partial z^{2}}f(z,s)=0.
\end{equation}
and
\begin{eqnarray}\label{heterogeneous_comb power-law f(y) final delta} 
-2\mathcal{D}_{x}z^{\alpha/2}\left.\frac{\partial}{\partial z}f(z,s)\right|_{z=0}=1.
\end{eqnarray}

\begin{figure}
\resizebox{1.0\textwidth}{!}{(a) \includegraphics{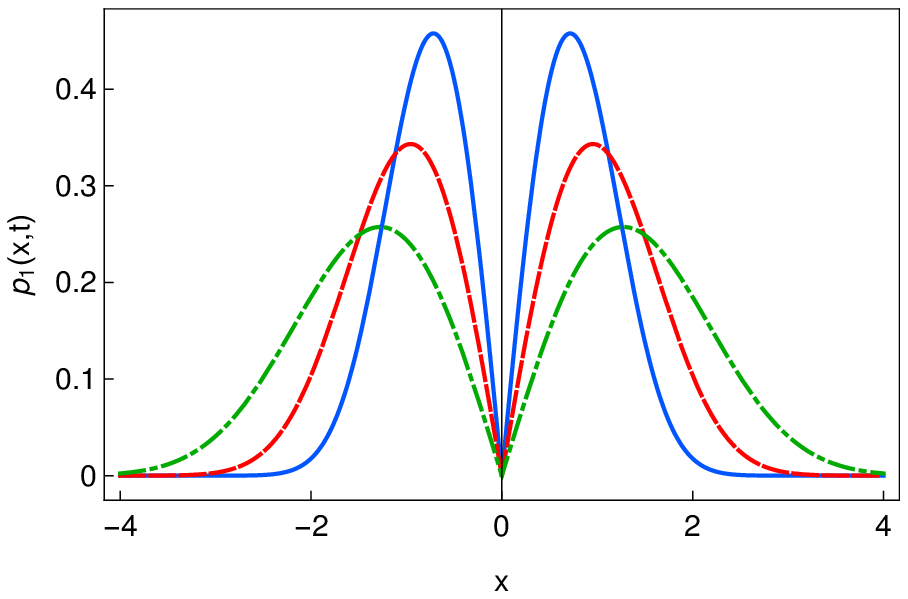} (b) \includegraphics{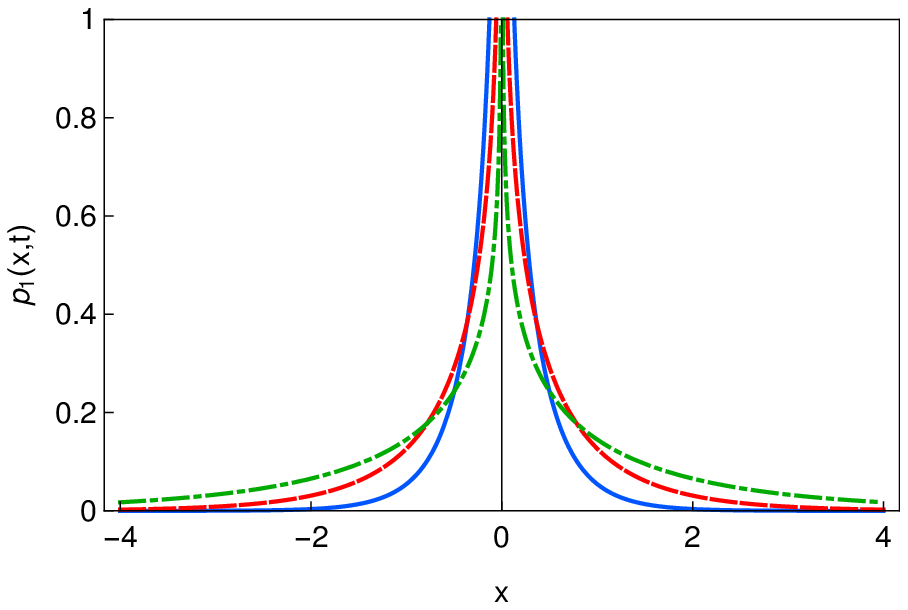}} 
\centering {\resizebox{0.5\textwidth}{!}{(c) \includegraphics{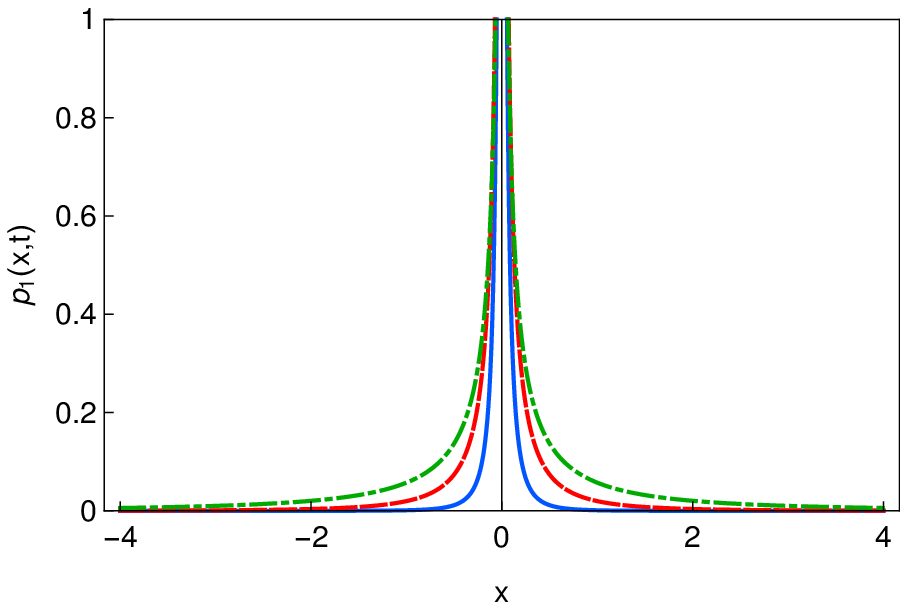}}}\caption {Graphical representation of the PDF (\ref{p1(x,t) final exact}) for $\mathcal{D}_{x}=1$, $\mathcal{D}_{y}=1$, (a) $\alpha=-2$ and $t=0.1$ (solid blue line), $t=1$ (red dashed line), $t=10$ (green dot-dashed line); (b) $\alpha=1/2$ and $t=0.1$ (solid blue line), $t=1$ (red dashed line), $t=10$ (green dot-dashed line); (c) $\alpha=5/4$ and $t=0.1$ (solid blue line), $t=1$ (red dashed line), $t=10$ (green dot-dashed line).}
\label{pdf plot}
\end{figure}

Eq.~(\ref{heterogeneous_comb power-law f(y) final homogeneous part}) is the Lommel equation, with the solution
\begin{eqnarray}\label{f(x,s) green sol K delta}
f(x,s)&=&\mathcal{C}(s)|x|^{(2-\alpha)/4}K_{\frac{1}{2}}\left(\frac{2s^{1/4}|x|^{(2-\alpha)/2}}{2-\alpha}\sqrt{\frac{2\mathcal{D}_{y}}{\mathcal{D}_{x}}}\right)\nonumber\\&=&\mathcal{C}(s)\frac{|x|^{(2-\alpha)/4}}{2}H_{0,2}^{2,0}\left[\left.\frac{s^{1/2}|x|^{2-\alpha}}{(2-\alpha)^{2}}\frac{2\mathcal{D}_{y}}{\mathcal{D}_{x}}\right|\left.\begin{array}{l} \\
(\frac{1}{4},1),(-\frac{1}{4},1)\end{array}\right.\right],
\end{eqnarray} 
where $\mathcal{C}(s)$ is a function which depends on $s$, $K_{\nu}(z)$ is the modified Bessel function (of the third kind) and $H_{p,q}^{m,n}(z)$ is the Fox $H$-function. The function $\mathcal{C}(s)$ is obtained by Eqs.~(\ref{heterogeneous_comb power-law f(y) final delta}) and (\ref{f(x,s) green sol K delta}), by using series representation of the modified Bessel function when $z\rightarrow0$ (\ref{K series}). Therefore, we find
\begin{eqnarray}\label{C(s)}
\mathcal{C}(s)=\frac{1}{\sqrt{4\pi}\mathcal{D}_{x}}\frac{2}{(2-\alpha)^{1/2}}\left(\frac{\mathcal{D}_{x}}{2\sqrt{\mathcal{D}_{y}}}\right)^{1/4}s^{-1/8}.
\end{eqnarray}
Thereafter, by using the inverse Laplace transform formula (\ref{H_laplace}) for the Fox $H$-function, and the properties (\ref{H_property}) and (\ref{H_property2}), for the reduced PDF, we finally obtain the solution
as follows
\begin{eqnarray}\label{p1(x,t) final exact}
p_{1}(x,t)=\frac{|x|^{-\alpha/2}}{\sqrt{4\pi\frac{\mathcal{D}_{x}}{2\sqrt{\mathcal{D}_{y}}}t^{1/2}}}H_{1,2}^{2,0}\left[\left.\frac{1}{(2-\alpha)^{2}}\frac{2\sqrt{\mathcal{D}_{y}}}{\mathcal{D}_{x}}\frac{|x|^{2-\alpha}}{t^{1/2}}\right|\left.\begin{array}{l} (3/4,1/2)\\
(1/2,1),(0,1)\end{array}\right.\right].\nonumber\\
\end{eqnarray}
Note that for $\alpha=0$ we recover the result for the classical comb
(\ref{diffusion eq on a comb}). Graphical representation of solution (\ref{p1(x,t) final exact}) is plotted in Fig.~\ref{pdf plot}.

From the reduced PDF we calculate the MSD 
\begin{eqnarray}\label{msd w}
\left\langle x^{2}(t)\right\rangle\simeq\frac{t^{\frac{1}{2-\alpha}}}{\Gamma\left(1+\frac{1}{2-\alpha}\right)}\sim t^{\beta}.
\end{eqnarray}
This corresponds to subdiffusion with the transport exponent $0<\beta<1/2$ for $\alpha<0$. For $0<\alpha<1$ subdiffusion is observed as well. Superdiffusion takes place for $1<\alpha<3/2$ since the transprt exponent is $1<\beta<2$. For the case with $3/2<\alpha<2$ one observes hyperdiffusion since $\beta>2$.

\subsection{Numerical analysis of MSD}

\begin{figure}
\centering {\resizebox{0.7\textwidth}{!}{\includegraphics{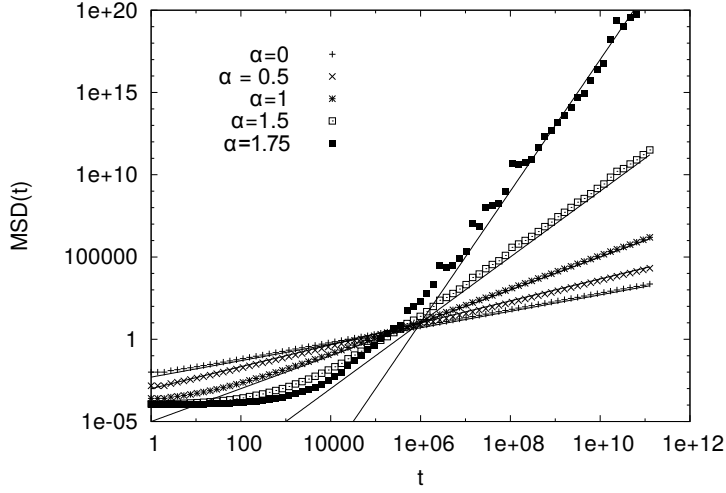} }}\caption {Graphical representation of numerically calculated ensemble averaged MSD $\langle x^2(t)\rangle$
for the CTRW model of Eq.~(\ref{heterogeneous_comb}) for different values of $\alpha$.
The straight lines represent the predicted power law behavior
$t^{\frac{1}{2-\alpha}}$.}
\label{figMSD}
\end{figure}

Furthermore, we give the numerically calculated ensemble averaged MSD $\langle x^2(t)\rangle$
for the CTRW model of Eq.~(\ref{heterogeneous_comb}). The waiting time in between successive jumps has PDF $\psi(t)\simeq t^{-3/2}$, which is the waiting time PDF in the fingers of the comb, with a cut-off at $t=1$. Each jump (the increment $\Delta x$ of the actual position $x$) is a sum of a deterministic part (drift) and a random part:
$$\Delta x=\delta t\cdot \alpha/4\cdot |x|^{\alpha-1}+\sqrt{\delta t}\cdot \xi\cdot |x|^{\alpha/2},$$
where, $\delta t$ is some fictitious time step of an integrator and determines the balance between drift and diffusion (in the simulations it is fixed to $\delta t=0.01$), $\xi$ is a gaussian random number with variance = 1. The drift is a consequence of the conversion of the Stratonovich stochastic differential equation Eq.~(\ref{LEq}) into an Ito stochastic differential equation which is then integrated using the Euler Maruyama scheme. Then the system is run where time is given by the sum of all waiting times. Graphical representation of the obtained MSD is given in Figure \ref{figMSD}. There exists good agreement in the long time limit with the analytical results (\ref{msd w}).

\section{Heterogeneous diffusion on a fractal grid}\label{sec3}

Considering heterogeneous diffusion on a fractal grid comb, we introduce an infinite number of backbones at positions $y$ which belong to the fractal set
$\mathcal{S}_{\nu}$ with fractal dimension $0<\nu<1$ \cite{fractal grid pre}. This situation is described by the diffusion equation
\begin{eqnarray}\label{heterogeneous_fractal grid} 
\frac{\partial}{\partial t}P(x,y,t)
&=&\mathcal{D}_{x}\sum_{l_{j}\in\mathcal{S}_{\nu}}\delta(y-l_{j})\frac{\partial}{\partial x}\left[|x|^{\alpha/2}\frac{\partial}{\partial x}\left(|x|^{\alpha/2}P(x,y,t)\right)\right]\nonumber\\&+&\mathcal{D}_{y}\frac{\partial^{2}}{\partial y^{2}}P(x,y,t).
\end{eqnarray}
Such kind of fractal structures is an idealization of more complex comb-like fractal networks in anisotropic porous media \cite{baklanov,maex}.

Integration of Eq.~(\ref{heterogeneous_fractal grid}) over $y$, and then the Laplace transform, yield
\begin{eqnarray}\label{heterogeneous_fractal grid p1 L} 
sp_{1}(x,s)-\delta(x)=\mathcal{D}_{x}\frac{\partial}{\partial x}\left[|x|^{\alpha/2}\frac{\partial}{\partial x}\left(|x|^{\alpha/2}\sum_{l_{j}\in\mathcal{S}_{\nu}}P(x,y=l_{j},s)\right)\right].
\end{eqnarray}
We present the PDF $P(x,y,t)$ in the Laplace space by means the same ansatz (\ref{ansatz})
\begin{eqnarray}\label{P general}
P(x,y,s)=g(x,s)\exp\left(-\sqrt{\frac{s}{\mathcal{D}_{y}}}|y|\right),
\end{eqnarray}
from where it follows $p_{1}(x,s)=2g(x,s)\sqrt{\frac{\mathcal{D}_{y}}{s}}$. The summation in Eq.~(\ref{heterogeneous_fractal grid}) is over the fractal set $\mathcal{S}_{\nu}$, which corresponds to integration over the fractal measure $\mu_{\nu}\sim l^{\nu}$, such that $\sum_{l_{j}\in\mathcal{S}_{\nu}}\delta(l-l_{j})\rightarrow\frac{1}{\Gamma(\nu)}t^{\nu-1}$ is the fractal density, and $d\mu_{\nu}=\frac{1}{\Gamma(\nu)}l^{\nu-1}dl$ \cite{tarasov}, from where it follows
\begin{eqnarray}\label{sum fractal grid p1} 
&&\sum_{l_j\in\mathcal{S}_{\nu}}P(x,y=l_j,s)=g(x,s)\frac{1}{\Gamma(\nu)}\int_{0}^{\infty}dl\,l^{\nu-1}\exp\left(-\sqrt{\frac{s}{\mathcal{D}_{y}}}l\right)\nonumber\\&&=g(x,s)\left(\frac{\mathcal{D}_{y}}{s}\right)^{\nu/2}=\frac{s^{(1-\nu)/2}}{2\mathcal{D}_{y}^{(1-\nu)/2}}p_{1}(x,s).
\end{eqnarray}
From Eqs.~(\ref{heterogeneous_fractal grid p1 L}) and (\ref{sum fractal grid p1}), we obtain the equation for the reduced PDF in the closed form
\begin{eqnarray}\label{heterogeneous_fractal grid p1 L2} 
sp_{1}(x,s)-\delta(x)=\frac{\mathcal{D}_{x}}{2\mathcal{D}_{y}^{(1-\nu)/2}}s^{(1-\nu)/2}\frac{\partial}{\partial x}\left[|x|^{\alpha/2}\frac{\partial}{\partial x}\left(|x|^{\alpha/2}p_{1}(x,s)\right)\right].
\end{eqnarray} 
Here we also take into account symmetrical property of the equation. Therefore we are considering the symmetrical PDF $p_1(x,s)=p_1(|x|,s)$ and substitution $f(x,s)=|x|^{\alpha/2}p_{1}(x,s)$, and by exchanging $|x|=z$, we find
\begin{eqnarray}\label{f diff eq} 
&&sz^{-\alpha/2}f(z,s)-\frac{\mathcal{D}_{x}s^{(1-\nu)/2}}{2\mathcal{D}_{y}^{(1-\nu)/2}}\frac{\alpha}{2}z^{\alpha/2-1}\frac{\partial}{\partial z}f(z,s)-\frac{\mathcal{D}_{x}s^{(1-\nu)/2}}{2\mathcal{D}_{y}^{(1-\nu)/2}}z^{\alpha/2}\frac{\partial^{2}}{\partial z^{2}}f(z,s)\nonumber\\&&-2\frac{\mathcal{D}_{x}s^{(1-\nu)/2}}{2\mathcal{D}_{y}^{(1-\nu)/2}}z^{\alpha/2}\frac{\partial}{\partial z}f(z,s)\delta(x) =\delta(x),
\end{eqnarray} 
from where one find the following system of differential equations:
\begin{eqnarray}\label{f diff eq1} 
\frac{\partial^{2}}{\partial z^{2}}f(z,s)+\frac{\alpha/2}{z}\frac{\partial}{\partial z}f(z,s)-\frac{2\mathcal{D}_{y}^{(1-\nu)/2}}{\mathcal{D}_{x}}s^{(1+\nu)/2}z^{-\alpha}f(z,s)=0,
\end{eqnarray} 
\begin{eqnarray}\label{f diff eq2} 
-2\frac{\mathcal{D}_{x}}{2\mathcal{D}_{y}^{(1-\nu)/2}}s^{(1-\nu)/2}\left.z^{\alpha/2}\frac{\partial}{\partial z}f(z,s)\right|_{z=0}=1.
\end{eqnarray} 
Therefore, for the final results we obtain
\begin{eqnarray}\label{grid p1(x,t) final exact}
p_{1}(x,t)&=&\frac{1}{\sqrt{4\pi\frac{\mathcal{D}_{x}}{2\mathcal{D}_{y}^{(1-\nu)/2}}t^{(1+\nu)/2}}}|x|^{-\alpha/2}\nonumber\\&\times&H_{1,2}^{2,0}\left[\left.\frac{1}{(2-\alpha)^{2}}\frac{2\mathcal{D}_{y}^{(1-\nu)/2}}{\mathcal{D}_{x}}\frac{|x|^{2-\alpha}}{t^{(1+\nu)/2}}\right|\left.\begin{array}{l} ((3-\nu)/4,(1+\nu)/2)\\
(1/2,1),(0,1)\end{array}\right.\right],\nonumber\\
\end{eqnarray} 
from where the MSD becomes
\begin{eqnarray}\label{grid msd w}
\left\langle x^{2}(t)\right\rangle\simeq\frac{t^{\frac{1+\nu}{2-\alpha}}}{\Gamma\left(1+\frac{1+\nu}{2-\alpha}\right)}.
\end{eqnarray}
Therefore, the diffusion is enhanced in comparison to the heterogeneous diffusion in a comb, as it is expected.

\section{Summary}\label{sec4}
The study is concerned with an inhomogeneous diffusivity of media in the comb-like geometry. We presented exact analytical results for various realizations of anomalous diffusion with a power-law-range-dependent diffusion coefficient $D(x)\sim |x|^{\alpha}$ with $\alpha<2$ along the main channels (backbones) on a comb (with one backbone) and a grid comb structures with a fractal backbone structure. For both cases of the comb and the fractal grid, the exact analytical solutions for the PDFs are obtained in the form of the Fox $H$-function and the mean square displacements are rigorously estimated as well. For the comb, the analytical form of the MSD $\langle x^2(t)\rangle\sim t^{1/(2-\alpha)}=t^{\beta}$, in Eq. (\ref{msd w}) reflects different realizations of anomalous diffusion, depending on the values of $\alpha$. These regimes are {\it (i)} subdiffusion with the transport exponent $0<\beta<1$ for $\alpha<1$; {\it (ii)} for $1<\alpha<3/2$ superdiffusion takes place with $1<\beta<2$; {\it (iii)} for the case with $3/2<\alpha<2$ one observes hyperdiffusion with the transport exponent $\beta>2$. Another important result is chaotic localization with the negative transport exponent $\beta<0$, when $\alpha>2$. In the case of the fractal grid, the fractal dimension of the backbone structure increases the transport exponent, which depends on the fractal dimension of the structure of the backbones, i.e., $\langle x^2(t)\rangle\sim t^{(1+\nu)/(2-\alpha)}$, where $0<\nu<1$ is the fractal dimension of the backbones structure. Here also, all regimes from subdiffusion to hyperdiffusion and localization take place as well.

\section*{Acknowledgments}
TS acknowledges support within DFG -- Deutsche Forschungsgemeinschaft project ``Random search processes, L\'evy flights, and random walks on complex networks". TS and AI thank the hospitality at the Max-Planck Institute for the Physics of Complex Systems in Dresden, Germany. AI was also supported by the Israel Science Foundation (ISF).

\appendix

\section{Fox $H$-function}

The Fox $H$-function is defined as the inverse Mellin transform for a set of gamma functions \cite{saxena book}
\begin{eqnarray}
H_{p,q}^{m,n}\left[z\left|\begin{array}{c l}
    (a_p,A_p)\\
    (b_q,B_q)
  \end{array}\right.\right]&=&H_{p,q}^{m,n}\left[z\left|\begin{array}{l}(a_1,A_1),\ldots,(a_p,A_p)\\
(b_1,B_1),\ldots,b_q,B_q)\end{array}\right.\right]\nonumber\\&=&\frac{1}{2\pi\imath}\int_{\Omega}ds\,\theta(s)z^{-s},
\label{H_integral}
\end{eqnarray}
where
\begin{eqnarray}\label{theta}
\theta(s)=\frac{\prod_{j=1}^{m}\Gamma(b_j+B_js)\prod_{j=1}^{n}\Gamma(1-a_j-A_js)}{
\prod_{j=m+1}^{q}\Gamma(1-b_j-B_js)\prod_{j=n+1}^{p}\Gamma(a_j+A_js)},
\end{eqnarray}
with $0\leq n\leq p$, $1\leq m\leq q$, $a_i,b_j \in C$, $A_i,B_j\in R^{+}$, $i=1,
\ldots,p$, and $j=1,\ldots,q$. The contour $\Omega$ starting at
$c-i\infty$ and ending at $c+i\infty$ separates the poles
of the function $\Gamma(b_j+B_js)$, $j=1,...,m$ from those of the
function $\Gamma(1-a_i-A_is)$, $i=1,...,n$.

The Mellin-cosine transform of Fox $H$-function is given by \cite{saxena book}
\begin{eqnarray}
\label{cosine H}
&&\int_{0}^{\infty}{d}\kappa\,\kappa^{\rho-1}\cos(\kappa x)H_{p,q}^{m,n}\left[a\kappa^{\delta}\left|
\begin{array}{l}(a_p,A_p)\\(b_q,B_q)\end{array}\right.\right]\nonumber\\&&=\frac{\pi}{x^\rho}H_{q+1,p+2}^{n+1,m}\left[\frac{x^\delta}{a}\left|
\begin{array}{l}(1-b_q,B_q),(\frac{1+\rho}{2},\frac{\delta}{2})\\(\rho,\delta),
(1-a_p,A_p),(\frac{1+\rho}{2},\frac{\delta}{2})\end{array}\right.\right].
\end{eqnarray}

The Mellin transform of the Fox $H$-function yieds
\begin{eqnarray}
\int_0^{\infty}dx\,x^{\xi-1}H_{p,q}^{m,n}\left[ax\left|\begin{array}{l}(a_p,A_p)\\
(b_q,B_q)\end{array}\right.\right]=a^{-\xi}\theta(\xi),
\label{integral of H}
\end{eqnarray}
where $\theta(\xi)$ is defined in Eqs.~(\ref{H_integral}) and (\ref{theta}).

The inverse Laplace transform of the Fox $H$-function reads \cite{saxena book}
\begin{eqnarray}\label{H_laplace}
\mathcal{L}^{-1}\left[s^{-\rho}H_{p,q}^{m,n}\left[as^{\sigma}\left|\begin{array}{l}(a_p,A_p)\\
(b_q,B_q)\end{array}\right.\right]\right]=t^{\rho-1}H_{p+1,q}^{m,n}\left[\frac{a}{t^{\sigma}}\left|\begin{array}{l}(a_p,A_p),(\rho,\sigma)\\
(b_q,B_q)\end{array}\right.\right],\nonumber\\
\end{eqnarray}

Fox $H$-function has the following properties \cite{saxena book}
\begin{eqnarray}\label{H_property}
H_{p,q}^{m,n}\left[z^{\delta}\left|\begin{array}{c l}
    (a_p,A_p)\\
    (b_q,B_q)
  \end{array}\right.\right]=\frac{1}{\delta}\cdot H_{p,q}^{m,n}\left[z\left|\begin{array}{c l}
    (a_p,A_p/\delta)\\
    (b_q,B_q/\delta)
  \end{array}\right.\right],
\end{eqnarray}
\begin{eqnarray}\label{H_property2}
z^{\sigma}H_{p,q}^{m,n}\left[z\left|\begin{array}{c l}
    (a_p,A_p)\\
    (b_q,B_q)
  \end{array}\right.\right]=H_{p,q}^{m,n}\left[z\left|\begin{array}{c l}
    (a_p+\sigma A_p,A_p)\\
    (b_q+\sigma B_q,B_q)
  \end{array}\right.\right].
\end{eqnarray}

\section{Lommel equation}\label{app 2}

The solution of the Lommel differential equation 
\begin{eqnarray}\label{Lommel}
u''(x)-c^{2}x^{2\zeta-2}u(x)=0
\end{eqnarray}
is given in terms of the Bessel functions \cite{book integrals}
\begin{eqnarray}\label{modified Bessel}
u(x)=\sqrt{x}Z_{\frac{1}{2\zeta}}\left(\imath\frac{c}{\zeta}x^{\zeta}\right).
\end{eqnarray}
The Bessel function $Z_{\frac{1}{2\zeta}}(x)$ is given by $Z_{\frac{1}{2\zeta}}(x)=C_{1}J_{\frac{1}{2\zeta}}(x)+C_{2}N_{\frac{1}{2\zeta}}(x)$, where $J_{\frac{1}{2\zeta}}(x)$ is the Bessel function of the first kind and $N_{\frac{1}{2\zeta}}(x)$ is the Bessel function of the second kind (Neumann function). In our case, the Bessel function is with imaginary argument, therefore the solution of the Lommel equation (\ref{Lommel}) is given in terms of the modified Bessel function (of the third kind) \cite{book integrals}
\begin{eqnarray}\label{Bessel third kind}
u(x)=\sqrt{x}K_{\frac{1}{2\zeta}}\left(\frac{c}{\zeta}x^{\zeta}\right),
\end{eqnarray}
which satisfies the zero boundary conditions at infinity. 

The modified Bessel function (of the third kind) $K_{\nu}(z)$ is a special case of the Fox $H$-function \cite{saxena book}
\begin{eqnarray}
H_{0,2}^{2,0}\left[\frac{z^{2}}{4}\left|\begin{array}{l} \\
(\frac{a+\nu}{2},1),(\frac{a-\nu}{2},1)\end{array}\right.\right]=2\left(\frac{z}{2}\right)^{a}K_{\nu}(z).
\label{HK relation}
\end{eqnarray}
Its series representation for $z\rightarrow0$ is given by
\begin{eqnarray}
K_{\nu}(z)&\simeq&\frac{\Gamma(\nu)}{2}\left(\frac{z}{2}\right)^{-\nu}\left[1+\frac{z^{2}}{4(1-\nu)}+\dots\right]\nonumber\\&+&\frac{\Gamma(-\nu)}{2}\left(\frac{z}{2}\right)^{\nu}\left[1+\frac{z^{2}}{4(\nu+1)}+\dots\right], \quad \nu\notin Z.
\label{K series}
\end{eqnarray}

\end{document}